\documentclass[final,5p,times,twocolumn]{elsarticle}

\usepackage{hyperref,subfigure}

\journal{Nuclear Instruments \& Methods in Physics Research, Section A}

\begin{document}

\begin{frontmatter}

\title{Recent results with radiation-tolerant TowerJazz 180~nm MALTA Sensors }

\author[a]{Matt~LeBlanc\corref{mycorrespondingauthor}}\ead{matt.leblanc@cern.ch}
\author[b]{Phil~Allport}
\author[a]{Igancio~Asensi}
\author[c]{Dumitru-Vlad~Berlea}
\author[d]{Daniela~Bortoletto}
\author[e]{Craig~Buttar}
\author[a]{Florian~Dachs}
\author[a]{Valerio~Dao}
\author[f]{Haluk~Denizli}
\author[a,g]{Dominik~Dobrijevic}
\author[a]{Leyre~Flores}
\author[a]{Andrea~Gabrielli}
\author[b]{Laura~Gonella}
\author[h]{Vicente Gonz\'alez }

\author[a]{Giuliano~Gustavino}
\author[f]{Kaan~Oyulmaz}
\author[a]{Heinz~Pernegger}
\author[a,i]{Francesco~Piro}
\author[a]{Petra~Riedler}
\author[j]{Heidi~Sandaker}
\author[a]{Carlos~Solans}
\author[a]{Walter~Snoeys}
\author[g]{Tomislav~Suligoj}
\author[a,j]{Milou~van~Rijnbach}
\author[a]{Abhishek~Sharma}
\author[a,h]{Marcos V\'azquez N\'u\~{n}ez}
\author[a,k]{Julian~Weick}
\author[c]{Steven~Worm}
\author[k]{Abdelhak~Zoubir}
\cortext[mycorrespondingauthor]{Corresponding author}

\address[a]{Experimental Physics Department, Organisation Européenne pour la Recherche Nucléaire (CERN),\\ F-01631 Prévessin Cedex, France -- CH-1211 Genève 23, Geneva, Switzerland}
\address[b]{University of Birmingham, United Kingdom}
\address[c]{Deutsches Elektronen-Synchrotron (DESY), Platanenallee 6, D-15738 Zeuthen, Germany}
\address[d]{University of Oxford, United Kingdom}
\address[e]{University of Glasgow, United Kingdom}
\address[f]{Bolu Abant Izzet Baysal University, Turkey}
\address[g]{University of Zagreb, Croatia}
\address[h]{University of Valencia, Spain}
\address[i]{\'Ecole polytechnique f\'ed\'erale de Lausanne, Lausanne, Switzerland}
\address[j]{University of Oslo, Norway}
\address[k]{Technische Universit\"{a}t Darmstadt, Germany}

\begin{abstract}
To achieve the physics goals of future colliders, it is necessary to develop novel, radiation-hard silicon sensors for their tracking detectors. We target the replacement of hybrid pixel detectors with Depleted Monolithic Active Pixel Sensors (DMAPS) that are radiation-hard, monolithic CMOS sensors. We have designed, manufactured and tested the MALTA series of sensors, which are DMAPS in the 180 nm TowerJazz CMOS imaging technology. MALTA have a pixel pitch well below current hybrid pixel detectors, high time resolution ($< 2$ ns) and excellent charge collection efficiency across pixel geometries. These sensors have a total silicon thickness of between 50--300 µm, implying reduced material budgets and multiple scattering rates for future detectors which utilize such technology. Furthermore, their monolithic design bypasses the costly stage of bump-bonding in hybrid sensors and can substantially reduce detector costs. This contribution presents the latest results from characterization studies of the MALTA2 sensors, including results demonstrating the radiation tolerance of these sensors.
\end{abstract}


\end{frontmatter}


\section{Introduction}

\noindent Future high-energy physics experiments target measurements of the Standard Model with an unprecedented level of precision. One key benchmark will be the measurement of major Higgs boson branching fractions to percent-level precision, which implies performant $b$-~\emph{vs.}~$c$-hadron identification and large datasets~\cite{Bambade:2019fyw,Bai:2021rdg}. To achieve such performance future tracking detectors must be designed with low material budgets and fast readouts. They will contain large surface areas of active sensors, and will be exposed to high radiation doses. These constraints motivate development of novel, radiation-hard silicon sensors that can affordably solve such problems.

Monolithic sensors, fabricated in commercial foundry processes, are a technology that offers potential solutions to several of these challenges~\cite{Aleksa:2649646}. Such sensors integrate the readout electronics and sensor into the same silicon wafer, eliminating the costly bump-bonding step of state-of-the-art hybrid pixel detectors currently used in high-energy physics experiments. The sensor capciatance for CMOS sensors can be made small ($< 5$~fF), offering a higher voltage signal despite reduced thickness of the sensor. This implies decreased power consumption requirements ($<1~\mu$W / pixel), which could permit a significant reduction of powering and cooling services for future detectors. Reduced material budgets result in less multiple scattering of charged particles, improving impact parameter and momentum resolution as well as increasing the overall tracking efficiency of the charged particles produced in collider events.

The MALTA series of depleted monolithic active-pixel sensors (DMAPS)~\cite{Pernegger:2274477,PERNEGGER2021164381,Dyndal_2020} was designed for potential applications in the ATLAS experiment~\cite{PERF-2007-01} at the high-luminosity LHC (HL-LHC)~\cite{ZurbanoFernandez:2020cco}. They are manufactured in the TowerJazz 180~nm CMOS imaging process, with additional process modifications~\cite{Snoeys:2017hjn,Munker_2019} that enhance the lateral electric field into pixel corners to increase tolerance to non-ionising energy loss (NIEL). MALTA2 is the latest full-scale prototype in this series, measuring $10.12 \times 20.2$~mm with a pixel pitch of $36.4$~$\mu$m. This sensor features a fast asynchronous readout capable of operating at 5~Gbps~\cite{Cardella:2691881}, sufficient for environments with hit rates on the order of 100~MHz/$\mathrm{cm}^{2}$. MALTA2 improves on the original MALTA sensor by cascoding the M3 transistor and enlarging the M4 transistor and CS capacitor~\cite{9764367}. These changes result in reduced noise tails compared to the original MALTA when operating at the same threshold and bias.

\section{Front-end characteristics \& irradiation}

\begin{figure}
  \center
  \includegraphics[width=0.45\textwidth]{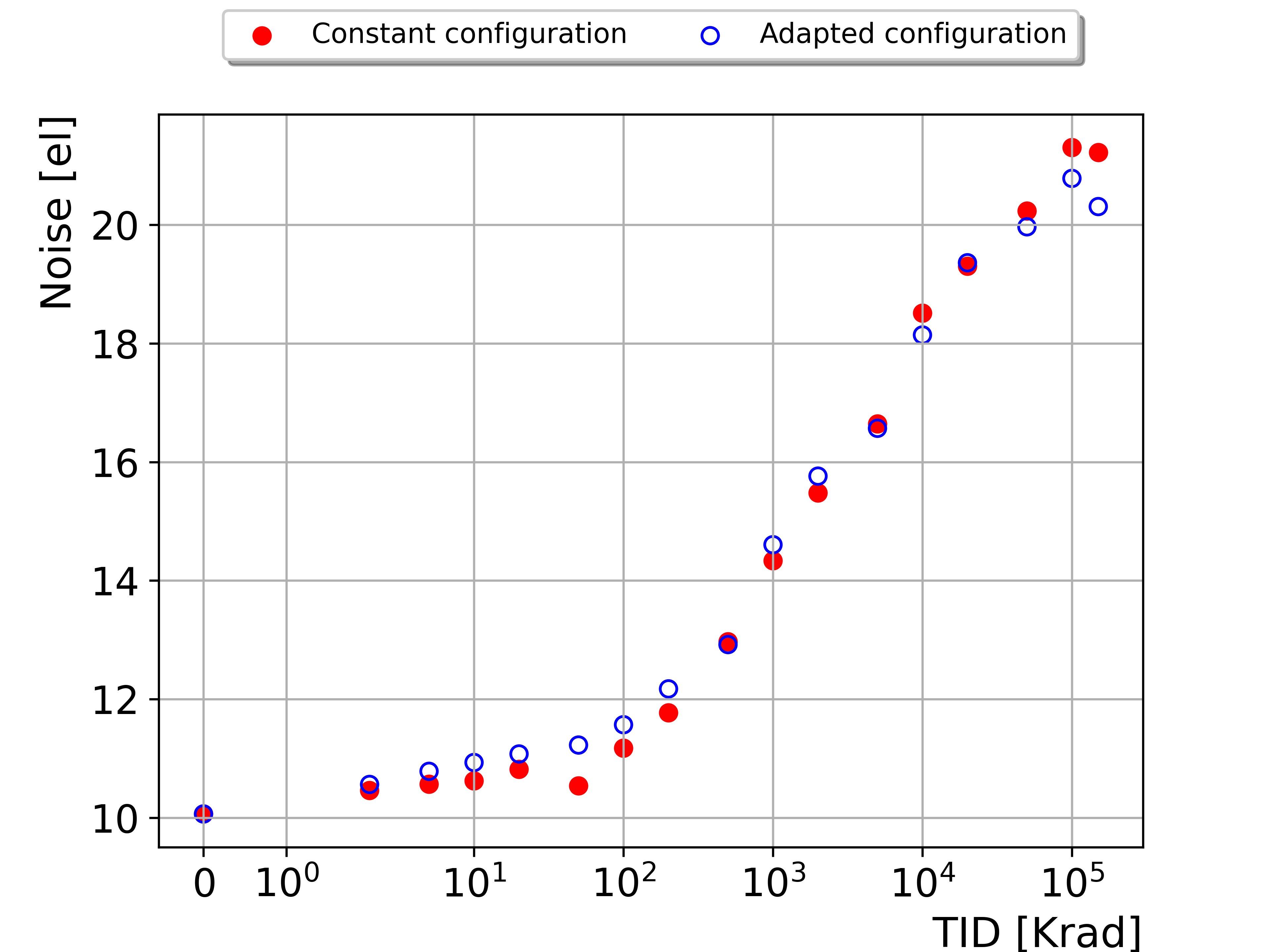}
  \caption{Equivalent noise charge \emph{vs.} total ionising dose for a MALTA2 sensor with high doping concentration in the $n$-blanket. The sensor was irradiated in steps up to 1~MRad, with (closed markers) no adjustments made to settings and (open markers) the threshold adjusted between each step to maintain a constant threshold of 500 $e^-$.}\label{fig:04}
\end{figure}

\noindent A series of studies from Ref.~\cite{9764367} were presented, which characterise the MALTA2 front-end electronics and demonstrate its radiation tolerance up to $3\cdot10^{15}~1~\mathrm{MeV}~n_{\mathrm{eq}}/\mathrm{cm}^2$ and $100$~MRad. The MALTA2 equivalent noise charge (ENC) is observed to increase monotonically \emph{vs.} TID when the front-end settings are adjusted to maintain a constant threshold of 100~$e^-$. A similar study of front-end irradiation to that of Ref.~\cite{9764367} was performed at a higher threshold (500 $e^-$) using a sensor with higher doping concentration in the $n$-type blanket implant, which produced a consistent picture of ENC development (figure~\ref{fig:04}).

The time-walk of the front-end was measured to be less than 25~ns for 90\% of signals from a ${}^{90}$Sr source. Charged particles produced by this source create MIP-like signals with an average charge deposition of roughly 1800 $e^-$, while signals with time-walk larger than 25~ns were observed for signals with charge depositions below 200 $e^-$. Such signals are likely produced by charge-sharing in pixel clusters.

 The time jitter of the front-end electronics was studied by injecting charge within a pixel using circuitry within the matrix digital readout. The arrival time of hits from the injected charge is compared to the timing of the charge injection trigger pulse transmitted to the chip, using a 3~ps binning TDC~\cite{picotdc}. The time jitter of the MALTA2 front-end electronics was measured to be 0.17~ns for injected charges above 1400 $e^-$, increasing to  4.7~ns at the nominal 100-$e^-$ threshold.

\section{Timing studies}

\noindent A campaign of test-beam measurements was performed with the 180 GeV proton beamline at the CERN Super Proton Synchrotron (SPS) during 2021, with the goal of characterising MALTA2 sensors in terms of their radiation tolerance and timing performance. A custom pixel telescope composed of six MALTA tracking \& triggering layers was used to study up to two MALTA2 devices under test (DUTs) at a time, hosted in a cold box. A scintillator located behind the telescope planes provides a timing reference for triggered signals.

Preliminary initial results from this campaign were presented, demonstrating the timing performance of two unirradiated MALTA2 sensors: one is a MALTA2 produced on a high-resistivity epitaxial layer, while the other is produced on a novel, thick, high-resisitivty (3-4 k$\Omega$) $p$-type Czochralski (Cz) substrate, respectively referred to as the 'epi.' and 'Cz' sensors in the following discussion. Both sensors have extra-deep $p$-well implants and are 100 $\mu$m thick. The epi. sensor has a low doping concentration of its $n$-type blanket, while the Cz sensor has a high doping concentration. Both sensors are operated  at -6 V substrate and $p$-well bias. The threshold of the Cz~sensor was measured to be to 170 $e^-$, while the threshold of the epi. sensor was measured to be 130 $e^-$.

\begin{figure}
  \center
  \subfigure[MALTA2, epi.]{\includegraphics[width=0.45\textwidth]{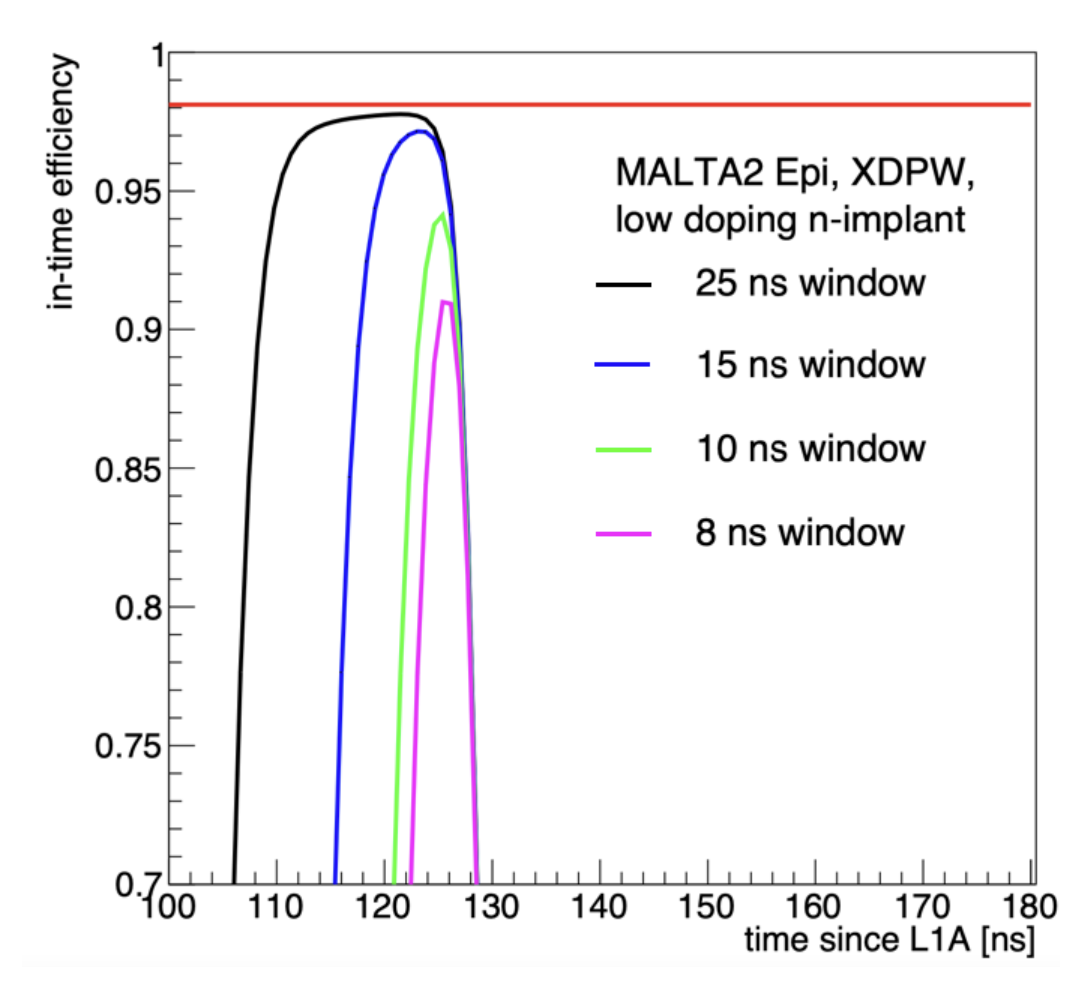}\label{fig:01:a}}
  \subfigure[MALTA2, Cz]{\includegraphics[width=0.45\textwidth]{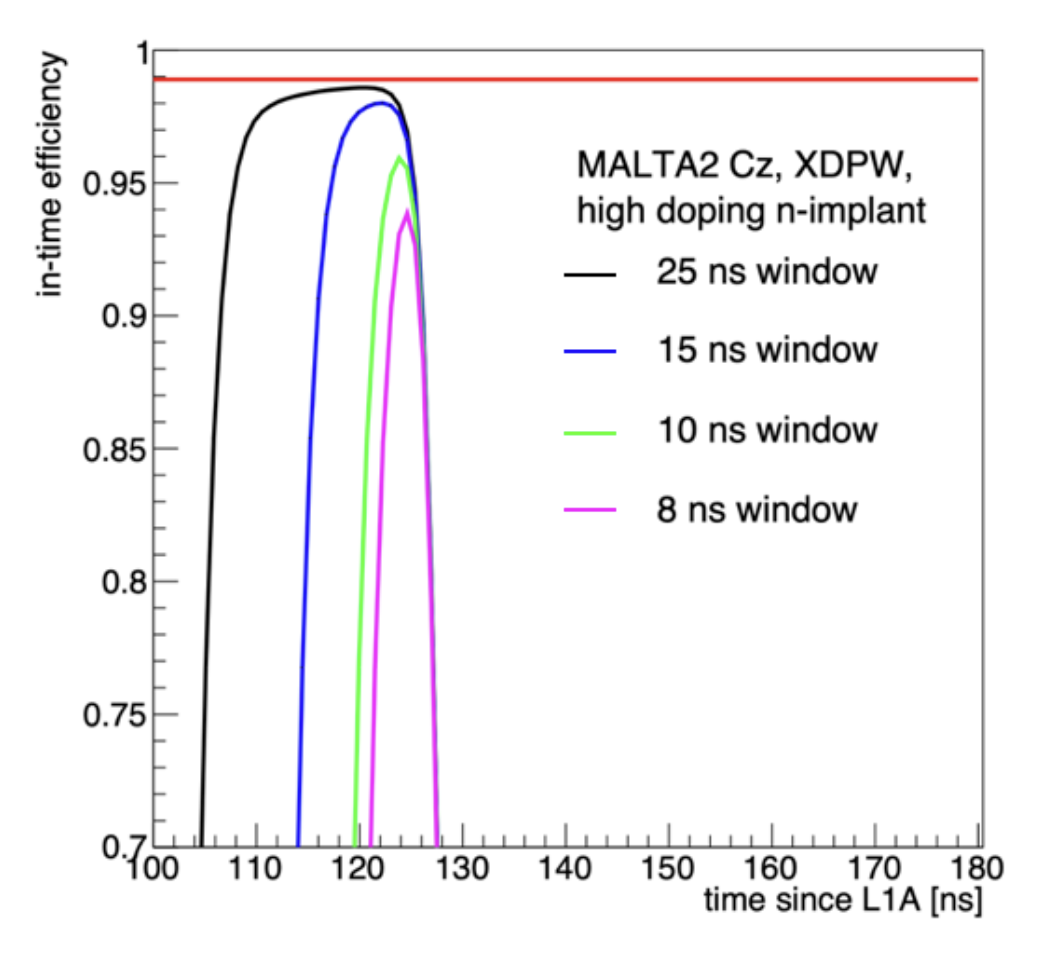}\label{fig:01:b}}
  \caption{In-time efficiency for MALTA2 sensors produced on (a) a high-resisitivity epitaxial layer and (b) a novel, thick high-resisitivty $p$-type Czochralski substrate. Both sensors are 100 $\mu$m thick and fabricated in the modified 180 nm TowerJazz process with extra-deep $p$-wells, with high doping of $n$-type blanket. Both sensors are operated at -6 V bias for both the substrate and $p$-well. The absolute value of the $x$-axis contains an arbitrary delay related to the signal processing pipeline, resulting in signals arriving after $\sim$130~ns. Measurements were performed with a 180~GeV proton beam at the CERN SPS during 2021.}\label{fig:01}
\end{figure}

The arrival time of the fastest hit in a pixel cluster was observed to increase slightly as a function of the distance of the hit in the pixel matrix from the front-end electronics. The arrival time delay also depends weakly on the front-end biasing group (columns 32 pixels wide) in the matrix where the hit is made. Correcting for these effects during offline reconstruction of the test-beam data results in an RMS time-of-arrival equal to 1.9~ns for the epi.~MALTA2, and 1.8~ns for the Cz~sensor. The in-time efficiency for both sensors was determined by integrating the time-of-arrival distributions with a sliding window algorithm, and is shown in figure~\ref{fig:01}. It was found to be above 98\% (90\%) for a 25~ns (8~ns) time window, suitable for applications at the HL-LHC and other proposed future collider facilities.

\begin{figure}
  \center
  \subfigure[MALTA2, epi.]{\includegraphics[width=0.45\textwidth]{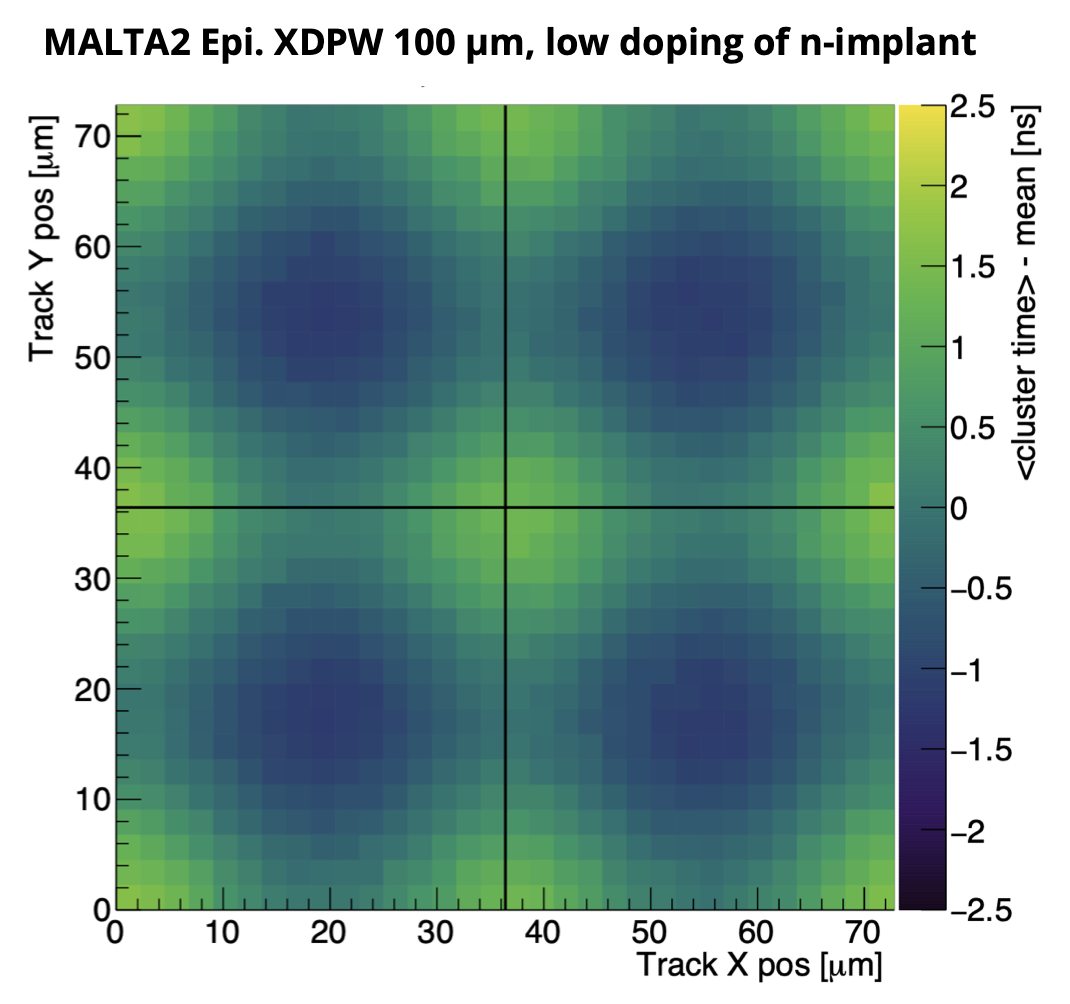}}
  \subfigure[MALTA2, Cz]{\includegraphics[width=0.45\textwidth]{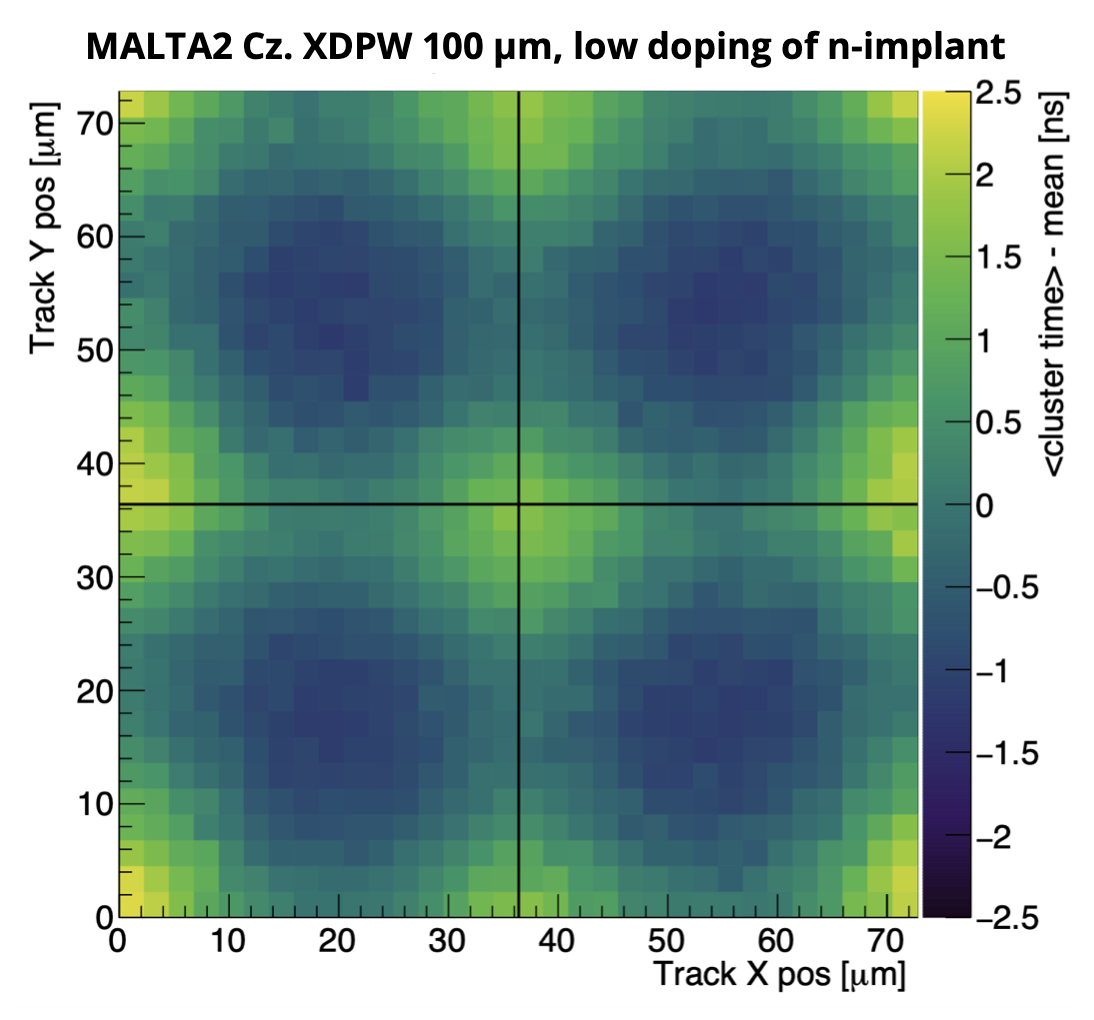}}
  \caption{In-pixel timing projected over a $2\times 2$ pixel matrix for MALTA2 sensors produced on (a) a high-resisitivity epitaxial layer and (b) a novel, thick high-resisitivty $p$-type Czochralski substrate. Both sensors are 100 $\mu$m thick and fabricated in the modified 180 nm TowerJazz process with extra-deep $p$-wells, with high doping of n-type blanket. Both sensors are operated at -6 V bias for both the substrate and $p$-well. Measurements were performed with a 180~GeV proton beam at the CERN SPS during 2021.}\label{fig:02}
\end{figure}

Figure~\ref{fig:02} shows, for the same two sensors, the difference between the arrival time of the leading hit in a pixel cluster and the average arrival time of signals from the entire pixel matrix. This information is projected over a $2\times 2$ pixel matrix, allowing charge-sharing effects to be studied. Signals with late arrival times are observed to originate from the corners of pixels: in such cases, charge-sharing results in a lower charge deposition per-pixel, and increased time-walk effects. More detailed simulations of charge-sharing effects in MALTA2 are being pursued.

\section{Multi-chip modules}

\begin{figure}
  \center
  \includegraphics[width=0.45\textwidth]{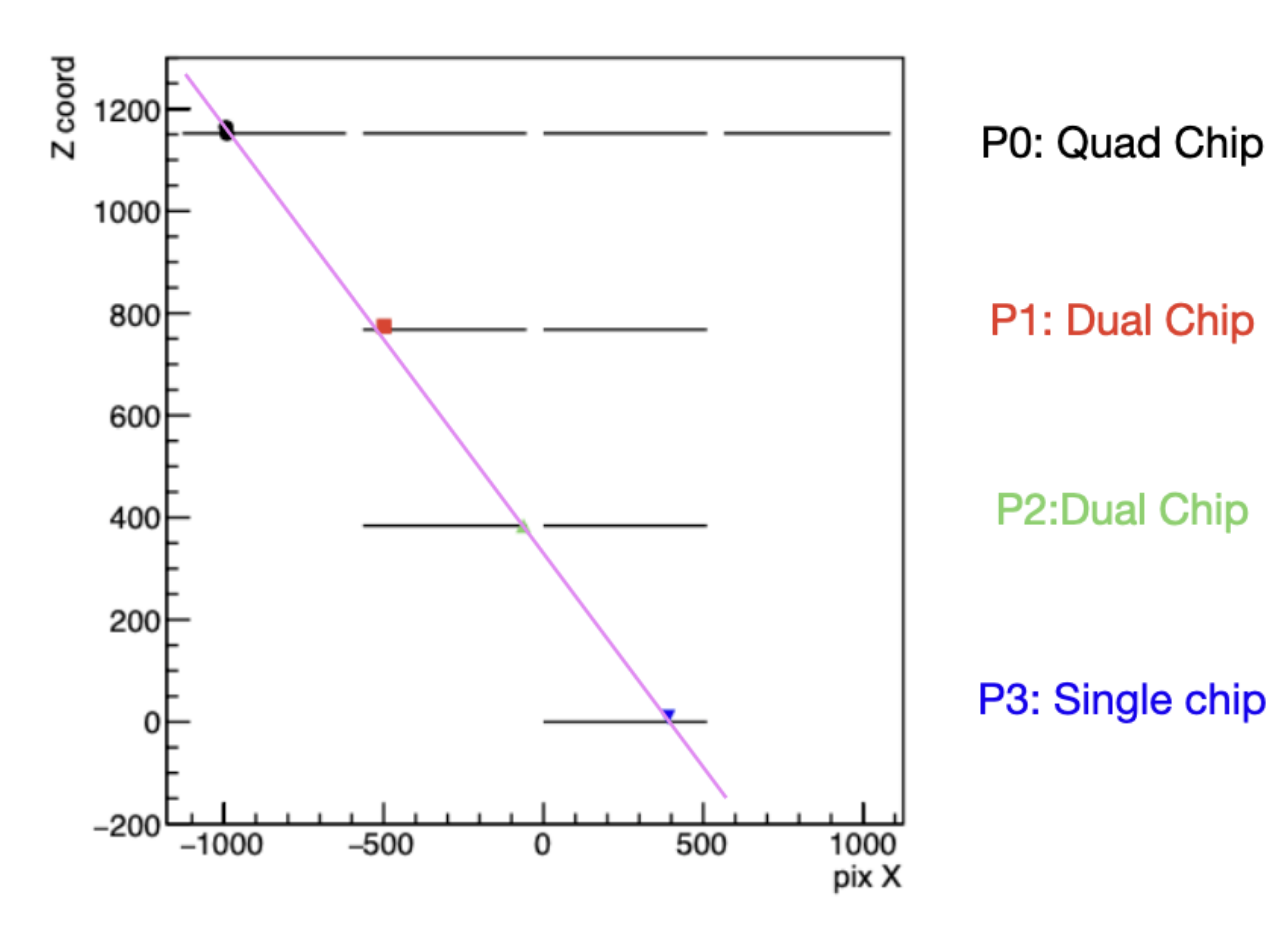}
  \caption{Event display of a cosmic-ray muon reconstructed using a lab-bench `mini-telescope' consisting of a quad-chip MALTA plane, a pair of dual-chip MALTA planes and a single-chip plane.}\label{fig:03}
\end{figure}

\noindent The production of multi-chip modules is ongoing for both MALTA and MALTA2 sensors. The functionality of dual-chip and quad-chip MALTA carrier boards was recently demonstrated in a lab-bench `mini-telescope' for cosmic ray data-taking (figure~\ref{fig:03}), where the multi-chip modules have been shown to significantly increase the acceptance of the apparatus. A four-chip flex PCB for MALTA2 sensors is being planned.

\section{Conclusion}

\noindent MALTA2 is the latest full-scale prototype monolithic sensor in the MALTA family, designed for use in experimental conditions similar to those of the HL-LHC. The front-end electronics of this sensor have been shown to be fully functional following irradiation up to $3\cdot10^{15}~1~\mathrm{MeV}~n_{\mathrm{eq}}/\mathrm{cm}^2$ and $100$~MRad~\cite{9764367}.

Initial studies of the MALTA2 timing performance have been performed. The time-walk of the front-end electronics was measured to be less than 25~ns for 90\% of signals from a ${}^{90}$Sr source, and the time jitter of the front-end electronics was found to be 0.16~ns for large input charges, increasing to 4.7~ns at the nominal 100-electron threshold. The pixel in-time efficiency was measured using a 180~GeV proton beam at the CERN SPS in summer 2021, and was found to be over 98\% (90\%) for a time window of 25~ns (8~ns) for MALTA2 sensors with high-resisitivty epitaxial layers or novel, thick $p$-type Czochralski~substrates manufactured with extra-deep $p$-well implants.

Initial results using multi-chip MALTA modules were presented, including quad- and dual-MALTA carrier boards and plans for a four-chip MALTA2 flex PCB.

\section*{Acknowledgements}

\noindent This project has received funding from the European Union’s Horizon 2020 Research and Innovation programme under Grant Agreement numbers 101004761 (AIDAinnova), 675587 (STREAM), and 654168 (IJS, Ljubljana, Slovenia).

\bibliography{mybibfile}

\end{document}